# Analysis of Cognitive Radio Scenes Based on Non-cooperative Game Theoretical Modelling

Ligia C. Cremene[1,3], D. Dumitrescu[2,3]

[1] Technical University of Cluj-Napoca, Memorandumului, 28, 400114, Romania
[2] Babes-Bolyai University, Kogalniceanu, 1, 400084, Cluj-Napoca, Romania
[3] Romanian Institute of Science and Technology, Saturn, 24, Cluj-Napoca, Romania
Ligia.Cremene@com.utcluj.ro

**Abstract.** A non-cooperative game theoretical approach for analyzing opportunistic spectrum access in cognitive radio environments is proposed. New concepts from computational game theory are applied to spectrum access analysis in order to extract rules of behaviour for an emerging environment. In order to assess opportunistic spectrum access scenarios of cognitive radios, two oligopoly game models are reformulated in terms of resource access: Cournot and Stackelberg games. Five cognitive radio scenes are analyzed: simultaneous access of unlicensed users ('commons regime') with symmetric and asymmetric costs, with and without bandwidth constraints, and sequential access ('licensed vs. unlicensed'). Several equilibrium concepts are studied as game solutions: Nash, Pareto, and the joint Nash-Pareto equilibrium. The latter captures a game situation where players are non-homogeneous users, exhibiting different types of rationality – Nash and Pareto. This enables a more realistic modelling of interactions on a cognitive radio scene. An evolutionary game equilibrium detection method is used. The Nash equilibrium indicates the maximum number of channels a cognitive radio may access without decreasing its payoff. The Pareto equilibrium describes a larger range of payoffs, capturing unbalanced as well as equitable solutions. The analysis of the Stackelberg modelling shows that payoffs are maximized for all users if the incumbents are Nash oriented and the new entrants are Pareto driven.

## 1 Introduction

Managing the development dynamics of wireless communications requires a systemic approach integrating user requirements and network resources. With the great extension of wireless connections, the complexity of interactions in the newly emerging environments reaches that of social interactions. This complexity renders conventional approaches to planning, implementation and regulation increasingly inadequate. This may be especially seen in the radio spectrum usage, which has manifested itself as a perceived a shortage of spectrum [1], [2]. This shortage is mainly due to inadequate command and control regulation – studies have shown up to 90% of the radio spectrum remains idle in any one geographical location [2], [3], [4].

Cognitive radios (CRs) are seen as the solution to the current low usage of the radio spectrum [5], [6], [7]. CRs have the potential to utilize the large amount of unused spectrum in an intelligent way while not interfering with other incumbent devices in frequency bands already licensed for specific uses [5]. Mechanisms based on frequency hopping have been widely used to enable wireless networks to use resources from the unlicensed spectrum without frequency planning [8]. In a CR environment, user nodes exhibit intelligent bevaviour, having the ability to observe, learn, and act in order to optimize their performance [3], [7]. Opportunistic spectrum access (OSA) dynamically obtains frequency assignments through sensing open spectral regions and adapting frequency selection in CRs. This ability has been under development within the technical community since 1999 and within the regulatory community since 2002 [1], [9].

Standard approaches to communications delivery need to be radically updated. This includes changes to radio regulation, business models, and economics. The application of distributed Computational Intelligence to communication devices – cognitive radios – is regarded as a key to the next revolution in communications delivery. This will enable intelligent local decisions to be made on network routing, spectrum and resource usage, based on interaction with other devices and the local environment [2]. Also, Computational Game Theory has emerged as an effective framework for the analysis and design of wireless networks, especially spectrum-aware networks [1], [14]. The suitability of using game theoretic concepts to model some problems in communications systems has been well argued during the last decade [10], [11], [12], [13], [14].

Seen as a promising approach for modelling and analyzing the interactions between CRs [10], [11], [14], [15], [16], Game Theory is a powerful tool in developing adaptive strategies in multiagent environments where multiple agents interact trying to achieve their own interests. In many cases their objectives conflict with each other. Cognitive radio environments are such environments – where the problem of dynamic spectrum sharing and issues of planning and decision-making have to be addressed. Eventually, adaptive strategies enable system components to learn a satisfactory policy (equilibrium) for action through repeated interaction with their environment [13]. Widely studied game models are exact potential games [13], [14], [16], [17]. The most frequently used steady-state concept is the Nash Equilibrium (NE) [10], [12], [14], [15], [16], [22]. Yet, there are other equilibria that may be relevant for real spectrum access scenarios.

In this paper, the problem of opportunistic spectrum access in cognitive radio environments is addressed, namely simultaneous and sequential access scenarios.

The paper is structured as follows: Section 2 introduces our general approach. Section 3 briefly describes the game equilibria that are studied as solution concepts in the analysis of CR scenes: Nash, Pareto, and joint Nash-Pareto equilibria. Section 4 presents opportunistic spectrum access scenarios that are modelled by reformulating the Cournot and Stackelberg oligopoly game models in terms of radio resource access. Numerical results are discussed. The conclusions are presented in Section 5.

## 2 General model

CR interactions are strategic interactions [12]: the utility of one CR (player) depends on the actions of all the other CRs on the scene. CR interactions may be modelled as games. A game model has the advantage of being simple and intuitive and it helps to identify and describe interaction situations that may occur. This allows the characterization of such situations and indicates the actions that would lead to them. Stable situations are described by game equilibria like: the standard Nash equilibrium, the non-standard Pareto equilibrium, and the new joint Nash-Pareto and Pareto-Nash equilibria. In our study, game equilibria are detected based on the generative relations presented in Section 3 and on an evolutionary search algorithm.

The interactions between CRs are modelled by reformulating two standard oligopoly game models – Cournot and Stackelberg – in terms of radio resource access. The type of game is important in that it may capture a certain model of interaction. For instance, the Cournot game is suitable for modelling open access of users – the so called ‚commons regime'; the Stackelberg game captures a sequential access model and that is why it is better suited for primary vs. secondary user access modelling.

The standard Cournot and Stackelberg economic competition models have been extensively used for pricing and spectrum trading problems [13], [15], [18]. Reformulations of economic game models in terms of radio resource access have also been studied, especially for power allocation and control [11], [15] [17], [19].

Our approach considers the number of simultaneously accessed channels by each CR [10], [16]. General, open, and opportunistic spectrum access scenarios are considered, as described in Section 4.

The main assumptions regarding CR interactions are: *(i)* CRs are spectrum-agile, have sensing and reconfiguration capabilities [5], [7], [9] and *(ii)* CRs do not know in advance what actions the other CRs will choose. We may assume that the CRs know the form of the other CRs' utility functions (if all the interacting CRs have the same objective, e.g., maximizing SINR). However, given the variability of channel conditions it is unlikely that a CR will know the precise values of other radios' utility functions [10]. That is why one-shot, non-cooperative games are considered here for capturing two and three CR interaction scenes.

Non-homogenous, self-regarding, players are assumed, exhibiting different types of rationality: Nash or Pareto. This approach embodies a new rationality paradigm in GT [20] that enables a more realistic modelling of the radio scene by capturing the heterogeneity of players. Within this paradigm CRs may have several approaches and biases towards different equilibrium concepts.

# 3 Evolutionary detection of game equilibria for cognitive radio scene analysis

Dynamic spectrum access scenes may be described as games between cognitive radios [10], [14], [19]: $n$ CRs attempting to access a frequency spectrum (e.g. the same set of channels at a certain time) interact on a radio scene. Several equilibrium concepts are studied as game solutions for opportunistic spectrum access: Nash, Pareto, and the joint Nash-Pareto equilibrium. A new approach for evolutionary detection of game equilibria based on generative relations [20], [21] is chosen.

A game may be defined as a system $G = (N, S_i, u_i, i = 1,…, n)$ where:
(i) $N$ represents the set of $n$ players, $N = \{1,…, n\}$,
(ii) for each player $i \in N$, $S_i$ represents the set of actions $S_i = \{s_{i1}, s_{i2}, …, s_{im}\}$;

$S = S_1 \times S_2 \times …S_n$ is the set of all possible game situations;

(iii) for each player $i \in N$, $u_i : S \to R$ represents the payoff function.

A strategy profile is a vector $s = (s_1,…, s_n)$, where $s_i \in S_i$ is a strategy (or action) of the player $i$.

By $(s_i, s_{-i}^*)$ we denote the strategy profile obtained from $s^*$ by replacing the strategy of player $i$ with $s_i$, i.e.

$$(s_i, s_{-i}^*) = (s_1^*, s_2^*,…, s_{i-1}^*, s_i, s_{i+1}^*,…, s_n^*).$$

## 3.1 Generative relation for Nash equilibrium

Game Theory provides a number of tools for analyzing games. The most frequently used steady-state concept is the Nash Equilibrium [22]. Informally, a strategy profile is a NE if no player can improve her payoff by unilateral deviation.

Formally, a strategy profile $x$ is a Nash equilibrium if and only if

$$u_i(x) \geq u_i(y_i, x_{-i}), \forall i \in N, y_i \in S_i .$$

A NE may be interpreted as a strategy from which no player has any incentive to deviate [12], [22].

A particular relation between strategy profiles can be used to describe a certain equilibrium.

Let $R$ be a binary relation on the strategy set $S$. Assume that $(x,y) \in R$ means that $x$ is better than $y$ with respect to a certain equilibrium concept $E$. We may also say that $x$ dominates $y$ with respect to $R$. A strategy $z$ is said to be non-dominated with respect to $R$ if and only if there is no strategy $x$ such that $x$ dominates $z$ ($(x,z) \in R$).

If the set of non-dominated strategies (with respect to $R$) equals the set $E$ then $R$ is said to be the *generative relation* of the equilibrium $E$ [20], [21].

A generative relation for the Nash equilibrium may be induced by a relative quality of two strategies.

Let us consider a binary quality measure [23] defined as

$$k(y, x) = card\{i : u_i(x_i, y_{-i}) > u_i(y), x_i \neq y_i\}$$

The quality $k(y,x)$ represents the number of players benefiting by switching from strategy $y$ to strategy $x$.

If $x^*$ is a Nash strategy then

$$k(x^*, x) = 0, \forall x \in S .$$

Using the relative quality measure $k$, a binary relation between strategies may be defined: $y$ is better than $x$ (with respect to NE), and we write $y \prec x$ or $(y, x) \in R$ if and only if less players benefit by switching from $y$ to $x$ than by switching from $x$ to y.

Therefore, we may write:

$$y \prec x \Leftrightarrow k(y, x) < k(x, y).$$

It may be proved [23] that the set of non-dominated strategies with respect to $R$ is a generative relation of the *Nash equilibrium*.

## 3.2 Generative relation for Pareto equilibrium

Let us consider two strategy profiles $x$ and $y$ from $S$.

**Definition.** The strategy profile *x* Pareto dominates the strategy profile *y* (and we write *x* < *P y*) if the payoff of each player using strategy *x* is greater or equal to the payoff associated to strategy *y*, and at least one payoff is strictly greater.

More formally we can write *x* < *P y* iff

$$u_i(x) \geq u_i(y), \text{ for each } i = 1, \ldots, n,$$

and there is some index *j* so that

$$u_j(x) > u_j(y).$$

The set of all non-dominated strategies with respect to the relation < *P* represents the *Pareto equilibrium* of the game [21].

### 3.3 Generative relation for joint Nash-Pareto equilibrium

The recently introduced Nash-Pareto and Pareto-Nash equilibrium concepts [21] capture game situations where players are biased towards different types of rationality – Nash or Pareto.

In an *n*-player game, let us consider that each player acts based on a certain type of rationality. Let us denote by $r_i$ the rationality type of the player $i = 1, \ldots, n$.

For opportunistic access on a CR scene, a two-player game, where $r_1$ = Nash and $r_2$ = Pareto, may be considered. The first player is biased towards the Nash equilibrium and the other one is Pareto-biased. In this case, a new type of equilibrium, called the *joint Nash-Pareto equilibrium*, may be considered [21].

Let us denote by $I_N$ the set of Nash-biased players (*N*-players) and by $I_P$ the set of Pareto-biased players (*P*-players). Therefore we have

$$I_N = \{i \in \{1,\ldots,n\}: r_i = Nash\},$$
$$I_P = \{j \in \{1,\ldots,n\}: r_j = Pareto\}.$$

An operator measuring the relative efficiency of profile strategies has been introduced [21].

$$E: S \times S \rightarrow N,$$

defined as

$$E(y, x) = card\{i \in I_N : u_i(x_i, y_{-i}) > u_i(y), x_i \neq y_i\}$$
$$+ card\{j \in I_P : u_j(y) < u_j(x), x \neq y\}.$$

*E( y, x)* measures the relative efficiency of the strategy profile *y* with respect to the strategy profile *x*. The relative efficiency enables us to define a generative relation for the joint Nash-Pareto (N-P) equilibrium.

Consider a relation <*NP* defined as *y* <*NP x* if and only if

$$E(y, x) < E(x, y).$$

The relation <*NP* is considered as the generative relation of the joint Nash-Pareto equilibrium [21].

### 3.4 Game equilibrium detection

An evolutionary technique for equilibrium detection is considered; it enables even the approximation of infinite equilibrium sets. Evolutionary multi-objective optimization algorithms [24] are efficient tools for evolving strategies based on a non-domination relation. A population of strategy profiles is evolved based on appropriate generative relations [20], [21] that allow the comparison of strategies. This comparison leads to a fitness assignment technique guiding the search of the evolutionary algorithm towards the game equilibrium. Therefore non-domination (with respect to a generative relation) is considered for fitness assignment purposes.

Numerical experiments aim the detection of either pure equilibria or a combination of equilibria paralleling cognitive radios interaction. An adaptation of the popular Non-dominated Sorting Genetic Algorithm NSGA2 [24] is considered. Thus, the complexity of the proposed approach is the complexity of NSGA2 [24].

# 4 Cournot and Stackelberg Game Modelling for Analyzing Cognitive Radio Scenes

In order to assess opportunistic spectrum access scenarios of cognitive radios, two oligopoly game models are considered. Cournot and Stackelberg economic competition models are reformulated in terms of radio resource access.

In the standard Cournot competition players (firms) simultaneously choose quantities to produce, not knowing the other players' actions [12]. In the Stackelberg oligopoly model players move sequentially – one player chooses its output, then the other player does so, knowing the output chosen by the first player [6]. The two models are suited for different scenarios, respectively: Cournot – for the 'commons regimes' (all users are unlicensed, and may simultaneously try to access a set of available channels) and Stackelberg – for 'licensed vs. unlicensed' scenarios (some users are already occupying certain channels – primary users (PUs), while the others are searching for channels to occupy – secondary users (SUs)).

In order to illustrate spectrum access situations, radio scenes with two and three CRs trying to access the same set of channels – simultaneously or sequentially – are analyzed. CRs' strategies and payoffs are represented two and three-dimensionally.

## 4.1 Cournot modelling of a CR scene – simultaneous spectrum access ('commons regime')

We consider a CR access scenario that can be modelled as a simple reformulation of the Cournot oligopoly game [10]. Suppose there are *n* cognitive radios attempting to access the radio scene (the same set of channels at a certain time). Each CR *i* is free to decide the number $c_i$, $c_i \geq 0$, of simultaneous channels the radio will access. The question is: how many simultaneous channels should each radio access in order to maximize its operation efficiency?

Based on the above scenario, a Cournot game can be reformulated according to Table 1:

**Table 1: Cournot game reformulation in terms of spectrum access**

| Players | The cognitive radios attempting to access a certain frequency band. |
|---|---|
| Actions | The strategy of each CR *i* is the number $c_i$ of simultaneously accessed channels; A strategy profile is a vector $c = (c_1, ..., c_n)$. |
| Payoffs | The difference between a function of goodput $P(C)c_i$ and the cost of accessing $c_i$ simultaneous channels $Kc_i$ (e.g. power consumption). |

A linear inverse demand function is considered – the number of non-interfered symbols *P(C)* is determined from the total number *C* of accessed channels (occupied bandwidth).

The demand function may be defined as:
$$P(C) = \begin{cases} B - C, & if C \leq B, \\ 0, & C > B, \end{cases} \quad (1)$$

where *B > 0* is the parameter of the inverse demand function, and

$C = \sum_{i=1}^{n} c_i$ is the total number of accessed channels.

The goodput for CR *i* is $P(C)c_i$. Radio *i*'s cost for supporting $c_i$ simultaneous channels is $C_i(c_i)$.
$$C_i(c_i) = Kc_i. \quad (2)$$

The payoff of CR *i* may then be written as:
$$u_i(c) = P(C)c_i - C_i(c_i).$$

In general, *P* decreases with the total number of implemented channels *C*, and $C_i$ increases with $c_i$ (more bandwidth implies more processing resources and more power consumption) [10].

If these effects are approximated by the linear functions (1), (2), the payoff function can be rewritten as:
$$u_i(c) = \left(B - \sum_{k=1}^{n} c_k\right)c_i - Kc_i,$$

where *B* is the number of available channels and *K* is the cost of accessing one channel.
The Nash equilibrium is considered as the solution of this game and is calculated as follows:
$$c_i^* = (B-K)/(n+1), \forall i \in N.$$
Pareto and Nash-Pareto equilibria are described by the generative relations [20,] [21] presented in Section 2.

### 4.2 Numerical experiments – simultaneous spectrum access, Cournot modelling

Radio scene assessment results are presented for the Cournot modelling with two and three CRs simultaneously trying to access a given set of channels. The emerging behaviour on the cognitive radio scene is captured by the detected equilibria (Fig. 1 and Fig. 2).

#### 4.2.1 *CR scene 1:* two CRs, simultaneous access, 24 channels available

The simulation parameters for the Cournot model are B = 24 and K = 3. For equilibrium detection the evolutionary technique from Section 2 is considered. A population of 100 strategies has been evolved using a rank-based fitness assignment technique. In all experiments the process converges in less than 20 generations.

Fig. 1 captures four types of equilibria: Nash, Pareto, Nash-Pareto, and Pareto-Nash. The four types of equilibria are detected in four separate runs.

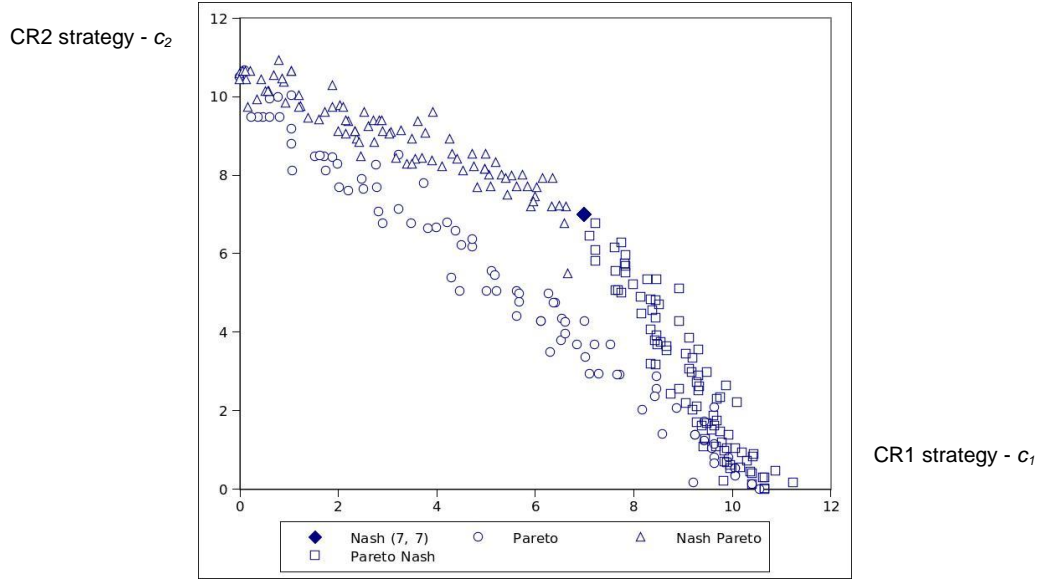

**Fig. 1.** Evolutionary detected strategies representing four types of equilibria: Nash (7,7), Pareto, Nash-Pareto, and Pareto-Nash. *CR scene 1:* two CRs, simultaneous access, B = 24, K = 3 (Cournot modelling).

NE of this scene corresponds to the situation where each of the two CRs activates 7 channels (from 24 available). The Pareto equilibrium (Fig. 1) describes a more unbalanced situation where the number of active channels lies in the range [0, 10.5] for each CR. Although each CR tries to maximize its utility, none of them can access more than half (12) of the available channels. Moreover, the sum of active channels on the Pareto front is less than the sum of active channels for the NE. The sum of simultaneously accessed channels is maximum at NE. Therefore, we may consider that this situation indicates an efficient use of the available spectrum, in terms of occupancy and fairness.

The joint Nash-Pareto equilibrium is achieved for the strategies on the N-P front. In some cases, the Nash-Pareto strategy enables the CR to access more channels than for the NE strategy. In the performed experiments (B = 24, K = 3; B = 10, K = 1; and B=100, K=1) the P-N equilibrium is symmetric to the N-P equilibrium with respect to the first bisecting line. It is interesting to notice that none of the N-P strategies actually reach NE.

Fig. 2 illustrates the payoffs of the two players, $u_1(c_1, c_2)$ and $u_2(c_1, c_2)$, corresponding to the four types of equilibria detected in a simultaneous access radio scene (Fig. 1).

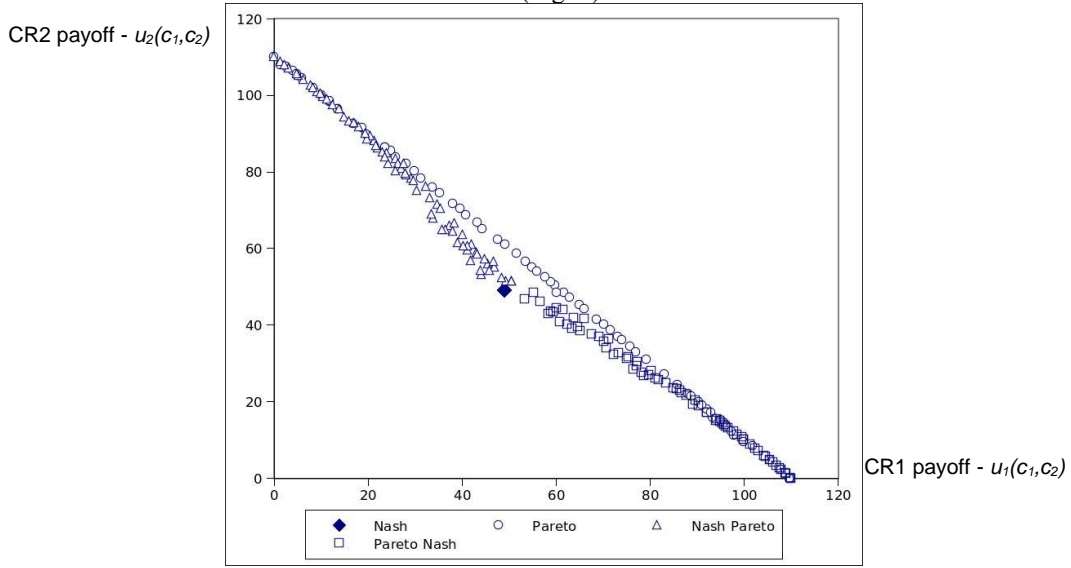

**Fig. 2.** Payoffs corresponding to the detected equilibria (Fig. 1) for CR scene 1: Nash (49, 49), Nash-Pareto, and Pareto-Nash.
*CR scene 1*: two CRs, simultaneous access, B = 24, K = 3 (Cournot modelling).

The symmetry of the joint equilibria payoffs (Pareto-Nash and Nash-Pareto) may be observed. The Pareto equilibrium (Fig. 2) describes a larger range of payoffs, capturing unbalanced as well as equitable solutions. The Pareto payoffs are in the range [0, 108] and their sum is always larger than the NE payoff (49, 49). For each strategy of the Nash-Pareto equilibrium the Pareto-player has a higher payoff. The Nash-player payoff is smaller in a Nash-Pareto situation than in a case where all the players play Nash (are Nash-biased).

Even if the N-P strategies allow the CRs to access more channels, the payoffs are smaller than for the Pareto strategies. This may be due to interference increasing with the number of simultaneously accessed channels.

#### 4.2.2 *CR scene 2:* **two CRs, simultaneous access, 100 channels available – scalability issue**

The evolutionary method for equilibrium detection is scalable with respect to the number of available frequency channels [25]. Fig. 3 illustrates the equilibrium strategies obtained for a larger set of channels (B = 100), with a unitary cost of accessing a channel (K = 1). The corresponding payoffs are captured in Fig. 4. Similar results are obtained by increasing the magnitude degree of B ($10^3$, $10^4$…) which is equivalent to having no bandwidth constraints.

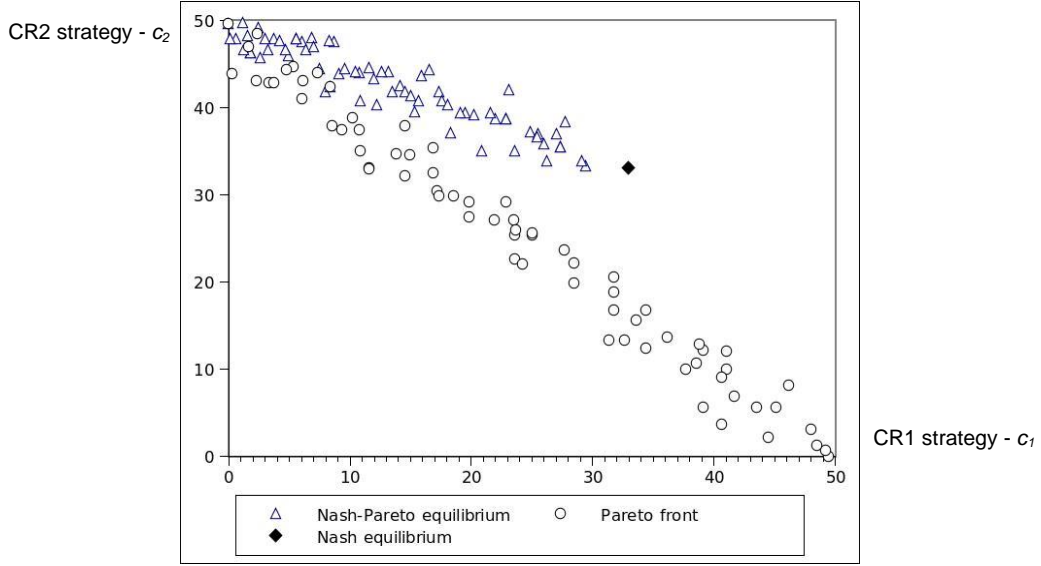

**Fig. 3.** Evolutionary detected strategies representing Nash (33,33), Pareto, and Nash-Pareto equilibria.
*CR scene 2:* two CRs, simultaneous access, B = 100, K = 1 (Cournot modelling).

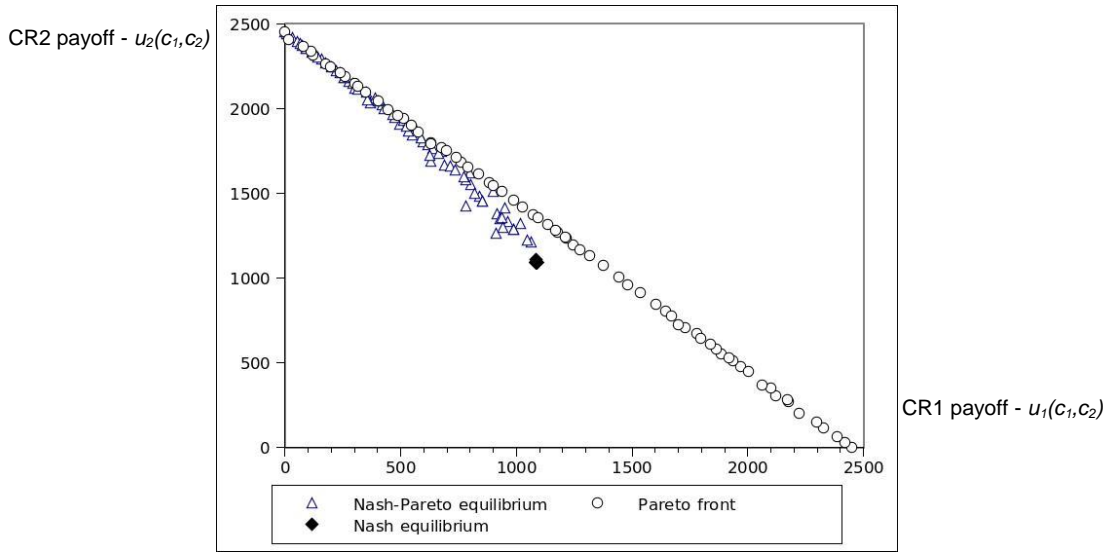

**Fig. 4.** Payoffs of the evolutionary detected equilibria for CR scene 2: Nash (1419, 1419), Pareto, and Nash-Pareto.
*CR scene 2:* two CRs, simultaneous access, B = 100, K = 1 (Cournot modelling).

**4.2.3** *CR scene 3:* **two CRs, simultaneous access, 24 channels, asymmetric costs**

An asymmetric scenario is considered where two CRs have different costs of accessing a channel. The Nash equilibrium is ($c_1^*$, $c_2^*$), where:

$$c_1^* = 1/3(B + K_2 - 2K_1),$$
$$c_2^* = 1/3(B + K_1 - 2K_2).$$

The situation corresponding to $K_1 = 1$ and $K_2 = 3$ is depicted in Fig. 5 (strategies) and Fig. 6 (payoffs).
The equilibrium payoff is higher for the CR having a lower access cost (in this case $CR_1$).

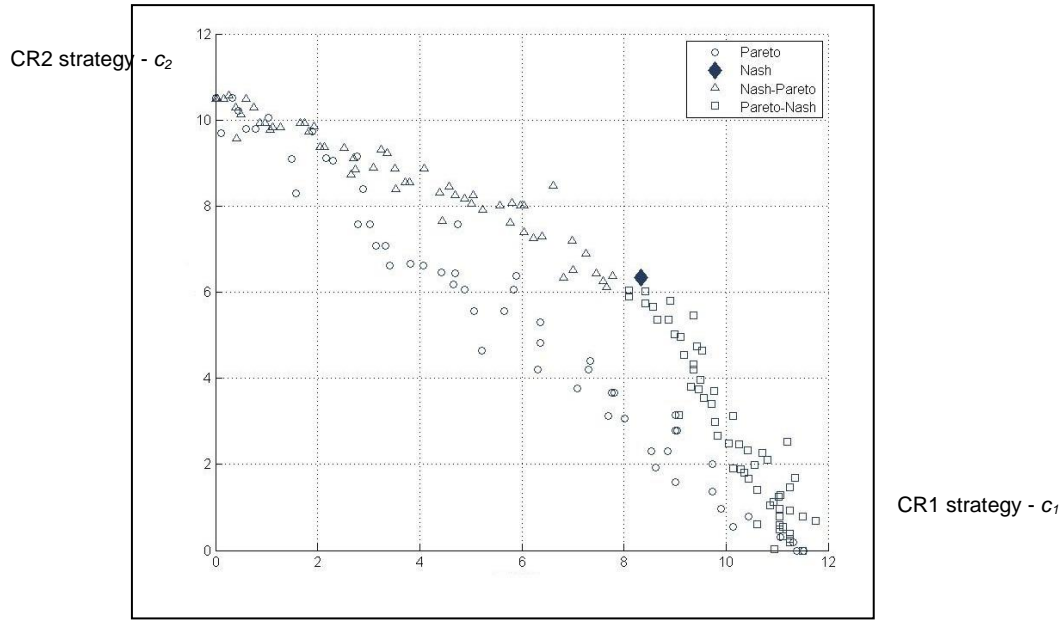

**Fig. 5.** Evolutionary detected strategies representing Nash (8.33, 6.33), Pareto, N-P, and P-N equilibria.
*CR scene 3:* two CRs, simultaneous access, different costs, B = 24, $K_1 = 1$, $K_2=3$ (asymmetric Cournot modelling).

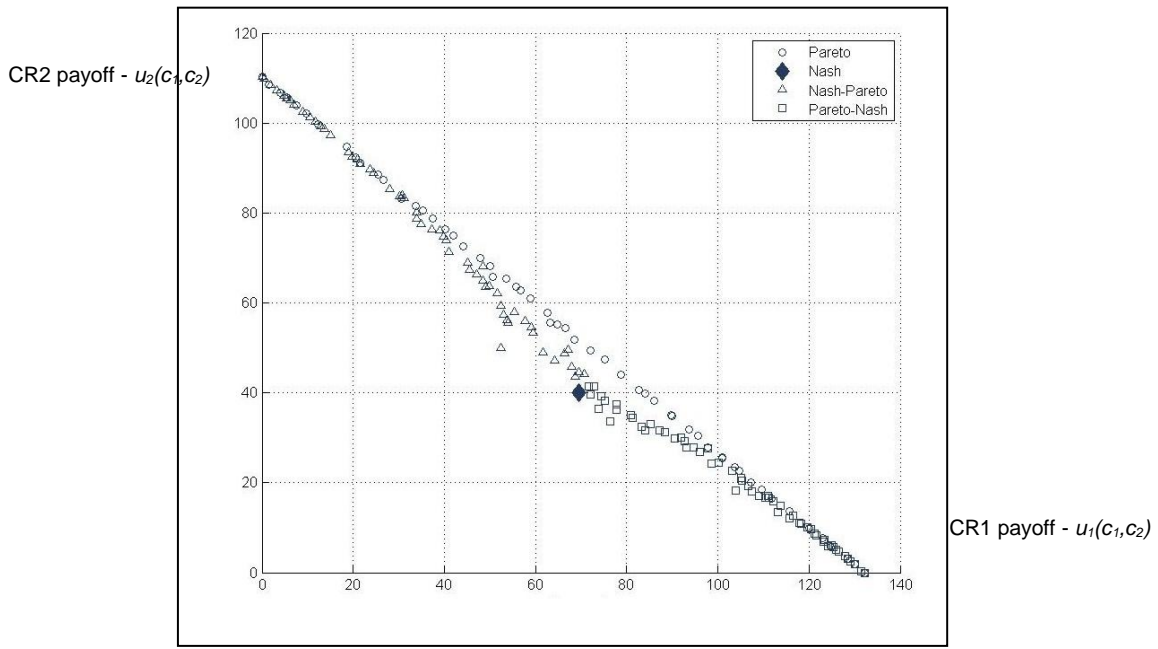

**Fig. 6.** Payoffs of the evolutionary detected strategies representing Nash (69.44, 40.11), Pareto, N-P, and P-N equilibria.
*CR scene 3:* two CRs, simultaneous access, different costs, B = 24, $K_1 = 1$, $K_2=3$ (asymmetric Cournot modelling).

### 4.2.4 *CR scene 4:* three CRs, simultaneous access, 24 channels available

In the 3-player game the variety of joint Nash-Pareto equilibrium situations increases combinatorially. For illustration we chose the Nash-Nash-Pareto equilibrium (Fig. 7 and Fig. 8).

The Nash equilibrium (Fig. 7) of the scene corresponds to the situation where each of the three CRs accesses 5 channels (from 24 available). The Pareto equilibrium describes a more unbalanced situation where the number of active channels lies in the range [0, 10.5] for each CR. Although each CR tries to maximize its utility, none of them can access more than approximately a third of the available channels. The Nash equilibrium, NE = (5.24, 5.24, 5.24), indicates the maximum number of channels a CR may access without decreasing its payoff. Hence, for this radio access scene, each CR may access a maximum of five channels.

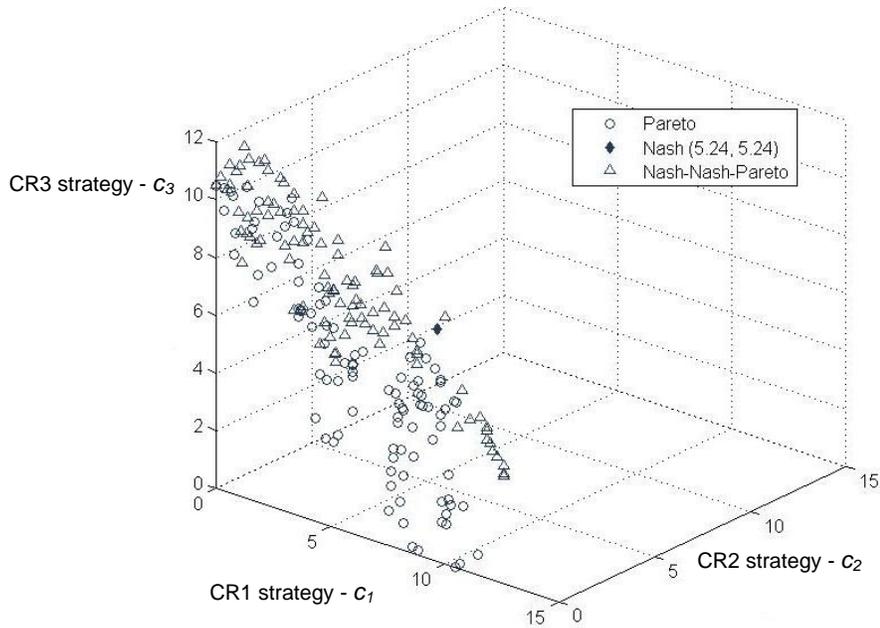

**Fig. 7.** Evolutionary detected strategies representing Nash (5.24, 5.24, 5.24), Pareto, and N-N-P equilibria. *CR scene 4:* three CRs, simultaneous access, B = 24, K = 3 (Cournot modelling).

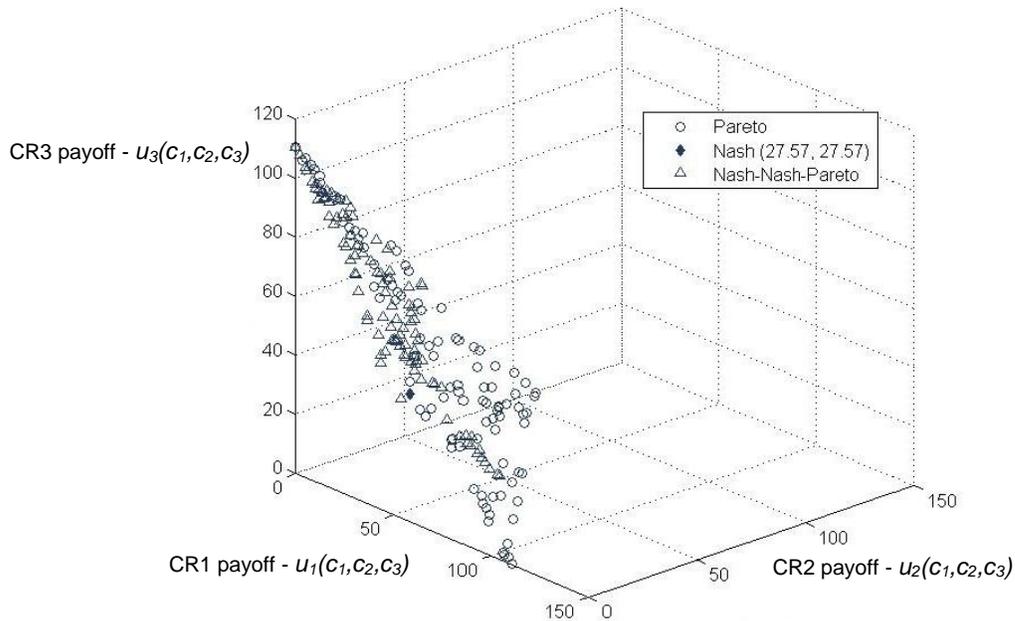

**Fig. 8.** Payoffs of the evolutionary detected equilibria for CR scene 4: Nash (27.57, 27.57, 27.57) Pareto, and N-N-P equilibria. *CR scene 4:* three CRs, simultaneous access, B = 24, K = 3 (Cournot modelling).

## 4.3 Stackelberg modelling of a CR scene – sequential spectrum access ('licensed vs. unlicensed')

The traditional spectrum access approach ensures co-existence of multiple systems by splitting the available spectrum into frequency bands and allocating them to licensed (primary) users. Dynamic spectrum access in CR environments improves the spectrum utilization by detecting whitespaces and making them available to unlicensed (secondary) users. This situation, where we have incumbent monopoly and new entrants, may well be modelled using a Stackelberg game model.

The Stackelberg game model accounts for priority of access of the primary users in the sense that PUs already occupy certain channels that SUs may happen to attempt accessing. Real 'do-no-harm' spectrum sharing is not possible in non-cooperative DSA with no additional information. So, when presuming OSA, we understand there may be situations where PUs and SUs may try to access the same channels, may compete, but the PUs always "choose" first and the SUs' actions will depend on the PUs' options (which is the essence of the Stackelberg game).

The same solution concepts as for the Cournot modelling are studied: Nash, Pareto, and the joint Nash-Pareto and Pareto-Nash equilibria.

The Stackelberg reformulation of the game in terms of spectrum access is captured by Table 2:

**Table 2: Stackelberg game reformulation in terms of spectrum access**

| Players | The cognitive radios – licensed and unlicensed (primary and secondary) users attempting to access a set of available channels. |
|---|---|
| Actions | The strategy of each CR $i$ is the number $c_i$ of simultaneously accessed channels; A strategy profile is a vector $c = (c_1,...,c_n)$. |
| Payoffs | The difference between a function of goodput $c_i\, P_d(C)$ and cost of accessing $c_i$ simultaneous channels $C_i(c_i)$ (e.g. power consumption). |

Using the same notations as for the Cournot modelling, the payoff function of CR $i$ may be defined as

$$u_i(c_1, c_2) = c_i P_d(c_1 + c_2) - C_i(c_i), \text{ for } i = 1,2,$$

considering $c_2 = b_2(c_1)$ as the output of the secondary user, for primary user's output $c_1$ [6].

A constant unit cost (K = 1) and a linear inverse demand function $P_d(C)$ are considered, with the same definition as for the Cournot model. The outcome of the equilibrium [12] is that CR 1 accesses $c_1^* = \frac{1}{2}(B-K)$ simultaneous channels and CR 2 accesses $c_2^* = b_2(c_1^*) = \frac{1}{4}(B-K)$ simultaneous channels.

### 4.4 Numerical experiments - sequential spectrum access, Stackelberg modelling

The simulation parameters for the Stackelberg model are $C_1(c_1) = C_2(c_2) = 3$ and $B = 24$. The evolutionary detected equilibria – Nash, Pareto, Nash-Pareto, and Pareto-Nash – are captured in Fig. 9. The corresponding payoffs are illustrated in Fig. 10.

#### 4.4.1 *CR scene 5:* two CRs (one PU, one SU), 24 channels available, sequential access

This situation is relevant for interference control in dynamic spectrum access scenarios between incumbents and new entrants. The Stackelberg modelling of such a scenario may provide particular answers to the fundamental question of autonomy versus regulation in heterogeneous radio environments [5], [9]. The analysis shows that payoffs are maximized for all users if the incumbents are Nash oriented and the new entrants are Pareto driven. This is actually consistent with a real-world scenario where incumbents want to keep their payoff unaffected (and play Nash) and the new entrants seek maximum payoff (and play Pareto).

We may notice that any strategy from the Pareto front is also a Nash-Pareto strategy. If the primary user plays Nash then the secondary user may maximize its payoff by choosing any strategy. If the secondary user plays Nash then the maximum payoff of the primary user is NE (10.5, 5.25).

Even if the SU can access less channels than in the Cournot case ($c_2$ = 5.25, which is less than 7 channels, NE = (7,7)), its maximum payoff remains unaffected, 108 (Fig.10). On the other hand, the PU's maximum payoff is half (55), even if it accesses more channels ($c_1$ = 10.5). For the Stackelberg formulation of the game, the NE payoff of the secondary user (Fig. 10) is less than in the Cournot case (25 instead of 49). The NE payoff of the primary user is slightly increased (55 instead of 49).

A unique Pareto-Nash equilibrium is detected and it overlaps the NE and also a Pareto strategy (Fig. 9). This indicates that NE is maintained even if the first player plays Pareto. In this case, NE is Pareto optimal. Therefore, the strategy where CR1 (PU) occupies 10 channels and CR2 (SU) occupies 5 channels is both stable (Nash) and optimal (Pareto).

Also, the sets of Nash-Pareto and Pareto equilibrium strategies overlap (Fig.9) (each N-P is also a Pareto strategy). This may indicate that Pareto optimality is maintained in most cases even if a CR plays Nash and the other one plays Pareto. As no new equilibrium is present, we may say that, for this particular instance of the game (B = 24, K = 3), heterogeneity of players does not affect existing game equilibria.

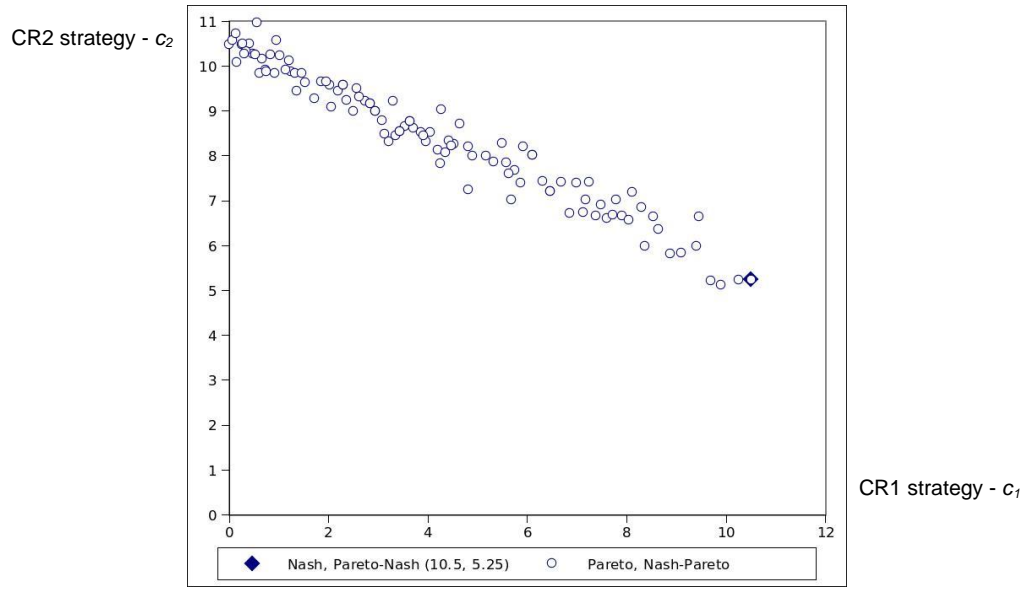

**Fig. 9.** Evolutionary detected equilibrium strategies: Nash and Pareto-Nash overlap (10.5, 5.25), Pareto and Nash-Pareto also overlap. *CR scene 5:* two CRs (one PU, one SU), $C_1(c_1) = C_2(c_2) = 3$, B = 24 (Stackelberg modelling).

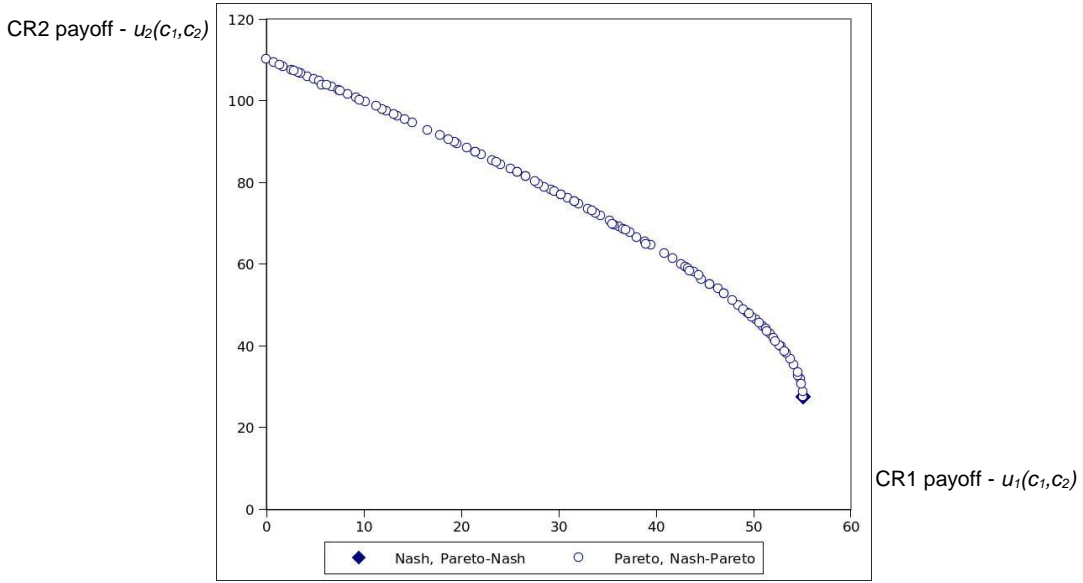

**Fig. 10.** Payoffs corresponding to the evolutionary detected equilibria: Nash and Pareto-Nash overlap (55,25), Pareto and Nash-Pareto overlap also. *CR scene 5:* two CRs (one PU, one SU), $C_1(c_1) = C_2(c_2) = 3$, $B = 24$ (Stackelberg modelling).

## 5  Conclusions

In order to investigate the relevance of certain game equilibrium concepts for emerging cognitive radio environments, two oligopoly game models are reformulated in terms of spectrum access: Cournot and Stackelberg. Two and three player opportunistic spectrum access scenarios are considered. Besides the standard Nash equilibrium new equilibrium concepts are investigated – the Pareto equilibrium and the joint Nash-Pareto equilibrium. The latter captures a game situation where players are non-homogeneous users, exhibiting different types of rationality: Nash and Pareto. This is a more realistic game formulation for interactions in CR environments, describing the natural asymmetry exiting among CRs.

Five CR scenes are analyzed: simultaneous access of unlicensed users ('commons regime') with symmetric and asymmetric costs, with and without bandwidth constraints, and sequential access ('licensed vs. unlicensed'). Numerical experiments indicate the relevance of the studied equilibria.

Nash equilibrium indicates the maximum number of channels a CR may access without decreasing its payoff. Also, the sum of simultaneously accessed channels is maximum at NE. Therefore, we may consider that this situation indicates an efficient use of the available spectrum, in terms of occupancy and fairness. The Pareto equilibrium indicates that, although each CR tries to maximize its utility, none of them can access more than half of the available channels without decreasing its payoff. In some cases, a joint Nash-Pareto strategy enables the CR to access more channels than for the NE strategy. For each strategy of the Nash-Pareto equilibrium the Pareto-player has a higher payoff.

The analysis of the Stackelberg modelling indicates that payoffs are maximized for all users if the incumbents are Nash oriented and the new entrants are Pareto driven. The observations may be especially relevant for designing new rules of behaviour for heterogeneous radio environments.

**Acknowledgments**

Thanks are due to Réka Nagy for her kind help in preparing the experiments.

This work was supported by the project "Development and support multidisciplinary postdoctoral programs in major technical areas of national strategy of the research - development - innovation" 4D-POSTDOC, contract no. POSDRU/89/1.5/S/52603, project co-funded from European Social Fund through Sectorial Operational Program Human Resources 2007-2013.

L. Cremene was partially supported by CNCSIS –UEFISCDI, Romania, PD grant 637/2010.